# Validation Measures in CMMI

Mahmoud Khraiwesh
Faculty of Science and Information Technology
Zarqa University
Zarqa - Jordan
mahmoud@zpu.edu.jo

*Abstract*—Validation is one of the software engineering disciplines that help build quality into software. The major objective of software validation process is to determine that the software performs its intended functions correctly and provide information about its quality and reliability.

This paper identifies general measures for the specific goals and its specific practices of Validation Process Area (PA) in Capability Maturity Model Integration (CMMI). CMMI is developed by Software Engineering Institute (SEI). CMMI is a framework for improvement and assessment of a software development process. CMMI needs a measurement program that is practical. The method we used to define the measures is to apply the Goal Question Metrics (GQM) paradigm to the specific goals and its specific practices of Validation Process Area in CMMI.

## I. INTRODUCTION

Validation is the process of evaluating a system or component during or at the end of the development process to determine whether it satisfies specified requirements. Validation means the software should do what the user really requires. Validation is a collection of analysis and testing activities across the full life cycle and complements the efforts of other quality engineering functions. Validation is a critical task in any engineering project [26].

Validation objective is to discover defects in a system and assess whether or not the system is useful and usable in operational situation. Validation should establishes confidence that the software is fit for its purpose and does what the user really requires [25].

Validation is concerned with checking that a computer program meets the requirements of the client [27]. Validation process, which is a part of the broader software process activities, plays a vital role in quality and profitability of the developed product [9].

Validation involves testing software or its specification at the end of the development effort to ensure that it meets its requirements (that it does what it is supposed to). Validation comprehensively analyzes and tests software to determine that it performs its intended functions correctly, to ensure that it performs no unintended functions, and to measure its quality and reliability. Validation is a system engineering discipline to evaluate software in a system context. Validation uses a structured approach to analyze and test the software against all system functions and against hardware, user, and other software interfaces [26].

Validation is an engineering process. The purpose of validation is to demonstrate that a product fulfills its intended use when placed in its intended environment. Validation is performed early and incrementally throughout the product life cycle [24].

Validation of requirements is a crucial activity in the software development process since it essentially determines the quality of the software product. Requirements validation is ensuring that the requirements adequately, consistently, and completely state the needs of the users [23].

Measurement is a mechanism for characterizing, evaluating, and predicting for various software processes and products [2]. The only way to improve any process is to measure specific attributes of the process, develop a set of meaningful metrics based on these attributes, and then use the metrics to provide indicators that will lead to strategy for improvement. Software measurement plays important role in understanding and controlling software development practices and products [16].

Measurement is the process by which numbers or symbols are assigned to attributes of entities in the real world in such a way as to characterize the attributes by clearly defined rules( and scales) [11]. Measurement is important for three basic activities: understanding, control and improvement [10]. Reasons for measuring are: to assess achievement of quality goals, to determine status with respect to plans, to gain understanding of processes, products, resources, and environments, to establish baselines for comparisons with future assessments and track improvement efforts [20]. The main measurement objective is to monitor software process performance [19].

Software measurement is currently in a phase in which terminology, principles and methods are still being defined and consolidated. We should not expect to find quantitative laws that are generally valid and applicable, and have the same precisions and accuracy as the laws of Physics, for instance. As a consequence, the identification of universally valid and applicable measures may be an ideal, long term research goal, which cannot be achieved in the near future, if at all [6]. Software engineering is not grounded in the basic quantitative laws of physics. Direct measure such as voltage, mass, velocity, or temperature, are uncommon in the software world. Because software measures and metrics are often indirect, they are open to debate [22]. There is a lack of an agreed-upon metrics validation framework [14]. The goal of software metrics is the improvement of the software process [8].

In the mid-1980s, the SEI initiated a study of ways of assessing the capabilities of software contractors. The outcome of this capability assessment was the Software Capability Maturity Model for Software (SW-CMM) [21]. The Software CMM was followed by other capability maturity models, including the People Capability Maturity Model (P- CMM) [7].

Organizations from industry, government, and the Software Engineering Institute (SEI) joined together to develop the CMMI Framework, a set of integrated CMMI models.

Two kinds of materials are contained in the CMMI model [1]:

1. Materials to help you evaluate the contents of your processes- information that is essential to your technical, support and managerial activities.
2. Materials to help you improve process performance-information that is used to increase the capability of your organization's activities.

Through the process of adopting CMMI, we aim to attain the following objectives: (1) to improve project management capability; (2) to enhance product quality; (3) to increase productivity and cost down; (4) to improve the capability of predicting the project budget and schedule; (5) to increase customer satisfaction [17].

With the adoption of CMMI comes the reward of software process improvement as well as product quality enhancement. It also defines a common language and a uniform standard for staff members to carry out daily tasks, and provides quantitative indicators for work performance and thereby further consolidates management [17].

CMMI model is quite comprehensive. It covers several bodies of knowledge and defined numerous process areas, specific and generic goals, specific and generic practices, as well as a lot of typical work products. It should be used to improve processes, increase productivity and raise competitiveness of an organization [28].

In CMMI-SW within each process area, there are one or more specific goals with specific practices and generic goals with generic practices. A specific goal applies to a process area and addresses the unique characteristics that describe what must be implemented to satisfy the process area. A specific practice is an activity that is considered important in achieving the associated specific goal. However, the CMMI recognizes that a specific practice is the goal rather than the way that goal is reached [25].

The Goal/Question/Metric (GQM) paradigm to process and metrics was developed by Basili and Weiss [4] as a technique for identifying meaningful metrics for any part of the software process. It has proven to be a particularly effective approach to selecting and implementing metrics.

This paper defines a general measure for the two specific goals and its five specific practices of Validation which is one of the PA in level 3 in CMMI-SW (Staged Representation) model. Measures will be compatible with the specific practices associated with the specific goal of Validation PA. The measures will be defined by applying the Goal Question Metrics (GQM) paradigm to the specific goal and its specific practices of Validation PA. The defined measures will help us to control and evaluate the software processes and products.

The remainder of the paper is organized as follows: section 2 describes related work in software measurement for the CMMI-SW, section 3 presents an overview of the CMMI-SW, section 4 presents an overview of the GQM, section 5 describes the application of the GQM to the CMMI-SW, section 6 describes the validity and reliability of the defined measures, and section 7 presents conclusions.

II. RELATED WORK

Many software measures activities have been proposed in the literature, some of them are [5] [13] [15] [18] [21]. The most related to our work are [5] [21] and [18].

Baumert and McWhinney [5] provide a set of indicators that are compatible with the measurement practices (one of the common features) described in the Capability Maturity Model for Software SW-CMM. These indicators cover thirteen categories. Not all categories occur at all maturity levels. They don't focus on a specific process. Their work was based on CMM not CMMI.

Paulk, Weber, Garcia, Crissis and Bush [21] provide a set of examples of measurements in measurement practices (one of the common features) of the Capability Maturity Model for Software (SW-CMM) in KPAs. A few examples related to requirements management KPA were provided. They don't focus on a specific process. Their work was based on CMM not CMMI. Loconsole [18] provided software measurements for implementation of Requirements Management KPA of the SW-CMM. Loconsole's work was based on CMM not CMMI.

This paper presents a set of general measures that are focused on a specific PA, Validation PA of the CMMI-SW. Measures are for the two specific goals and its five specific practices of Validation PA.

III. OVERVIEW OF THE CMMI-SW

The CMMI-SW (Staged Representation) is composed of five maturity levels: Initial, Managed, Defined, Quantitatively Managed and Optimizing. Figure1 shows the five maturity levels. Each maturity level is composed of several process areas with the exception of Level1 [24].

In CMMI-SW within each process area, there is one or more specific goals with specific practices and generic goals with generic practices. Generic goals are associated with the institutionalization of good practice, called "generic" because the same goal statement appears in multiple process areas as shown in figure 2. A specific goal applies to a process area and addresses the unique characteristics that describe what must be implemented to satisfy the process area. A specific practice is an activity that is considered important in achieving the associated specific goal [24].

The purpose of Validation is to demonstrate that a product or product component fulfills its intended use when placed in its intended environment [24].

The specific goals associated with Validation process area and the specific practices associated with each specific goal:

1- Prepare for validation: preparation for validation is conducted.
  1.1 Select product for validation.
  1.2 Establish the validation environment.
  1.3 Establish validation procedures and criteria.
2- Validate product or product components: the product or product-components are validated to ensure that they are suitable for use in their intended operating environment.
  2.1 Perform validation.
  2.2 Analyze validation results.

## IV. OVERVIEW OF THE GQM

The Goal/Question/Metric (GQM) paradigm is a method for helping an organization to focus the measurement program on their goals. It states that an organization should have specific goals in mind before data are collected [2]. The more mature your process, the more that is visible and therefore measurable. GQM does not specify concrete goals. It is rather a structure for defining goals and refining them into a set of quantifiable questions, these questions imply a specific set of metrics and data to be collected in order to achieve these goals.

The GQM paradigm consists of three steps:

1. Specify a set of goals based on the needs of the organization and its projects. Determine what should be improved or learned. The process of goal definition is supported by templates. By using these templates it is possible to define the goals in terms of purpose, perspective, and environment. Measurement goals should be defined in an understandable way and should be clearly structured. For this purpose, templates are available that support the definition of measurement goals by specifying purpose (what object and why), viewpoint (what aspect and who), and context characteristics [3].

A goal definition template [4] can be used to define each measurement goal. This template is illustrated in table 1, and takes the form:

Analyze {the name of activity or attribute to be measured} for the purpose of {the overall objective of the analysis} with respect to {the aspect of the activity or attribute that is considered} from the viewpoint of {the people who have an interest in the measurement} in the context of {the environment in which the measurement takes place}.

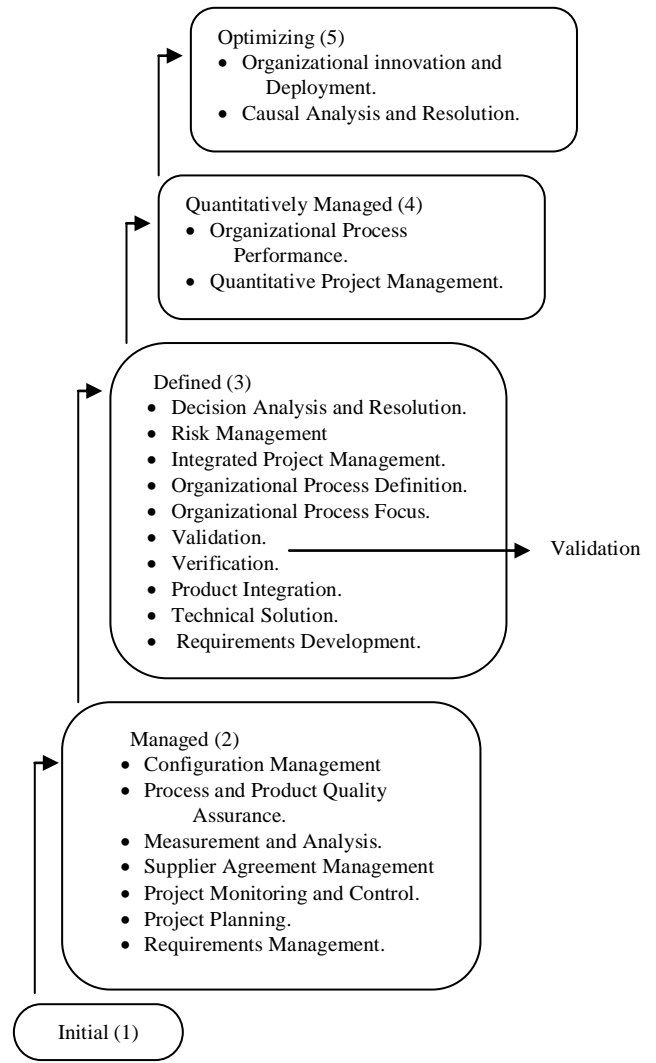

Figure 1. five levels with PA's in CMMI.

TABLE 1. GQM GOAL DEFINITION TEMPLATE.

| Analyze | the object under measurement |
|---|---|
| For the purpose of | understanding, controlling, or improving the object |
| With respect to | the quality focus of the object that the measurement focuses on |
| From the viewpoint of | the people that measure the object |
| In the context of | the environment in which measurement takes place |

2. Generate a set of quantifiable questions. Business goals are translated into operational statements with a measurement focus. Basili and Rombach [2] provide different sets of guidelines to classify questions as product-related or process-related.

3. Define a set of metrics that provide the quantitative information needed to answer the quantifiable questions. In this step, the metrics suitable to provide information to answer the questions are identified and related to each question. Several metrics might be generated from a single goal. Several measurements may be needed to answer a single question. Likewise, a single measurement may apply to more than one question.

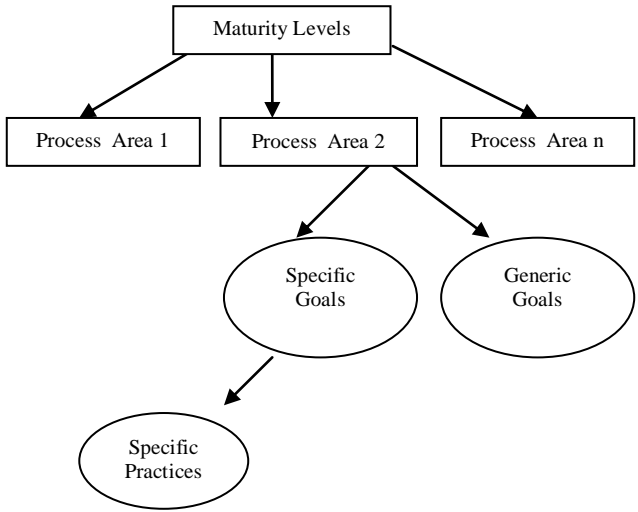

Figure 2. specific and generic goals

## V. APPLYING GQM TO THE CMMI-SW

The CMMI-SW defines two specific goals for Validation PA. There are 5 specific practices related to the specific goals. We deal with the specific practices as goals. We will apply the GQM on the five specific practices.

The five specific practices associated with Validation process area are:

1. Select product for validation: Select products and product components to be validated and the validation methods that will be used for each.

2. Establish the validation environment: Establish and maintain the environment needed to support validation.

3. Establish validation procedures and criteria: Establish and maintain procedures and criteria for validation.

4. Perform validation: Perform validation on the selected products and product components.

5. Analyze validation results: Analyze the results of the validation activities.

These five specific practices can be used for the first step of the GQM. The specific practices can be redefined by using the template in table1. The second step in the GQM paradigm is to generate a set of quantifiable questions. The third step of the GQM is to define a set of metrics that provide the quantitative information necessary to answer the questions. The sub practices and the typical work products which are found in each specific practice are take in consideration when we identify the set of measures.

A set of questions and measures is presented in the following tables, table2 through table6, each table related to one specific practice. There are overlaps among the questions and among the measures. The same measure can be used to give information to answer different questions.

### A. *Measures for specific practice 1.*

Select product for validation: Select products and product components to be validated and the validation methods that will be used for each.

A set of questions and measures is presented in the following table related to specific practice 1.

TABLE 2. SET OF QUESTIONS AND MEASURES FOR SPECIFIC PRACTICE 1.

| | Questions | Measures |
|---|---|---|
| Q1 | Do you select products and product components to be validated? | • Selecting products and product components to be validated.<br>• # Products to be validated.<br>• # Product components for each product to be validated.<br><br>(# means number of) |
| Q2 | Do you select the operational behavior for products and product components? | • Selecting of operational behavior for products and product components. |
| Q3 | Do you select user manual, user interfaces, maintenance services, and training materials to be validated? | • Selecting of user manual, user interfaces, maintenance services, and training materials to be validated. |
| Q4 | Do you identify the validation constraints for each product or product components? | • Identifying the validation constraints for each product or product components. |
| Q5 | Do you select the validation method for each product or product components?<br><br>(validation method includes: discussion with users, prototyping, functional demonstration testing by stakeholders, or pilots of training materials) | • Selecting the validation method for each product or product components. |
| Q6 | Do you select the validation method early in the life of the project? | • Selecting the validation method early in the life of the project. |
| Q7 | Do you share the stakeholders when selecting the validation method? | • Sharing the stakeholders when selecting the validation method. |
| Q8 | Do you determine the categories of user need to be validated? | • Determining the categories of user needs to be validated. (categories: operational, user manuals, or training) |

### B. *Measures for specific practice 2.*

Establish the validation environment: Establish and maintain the environment needed to support validation.

A set of questions and measures is presented in the following table related to specific practice 2.

TABLE 3. SET OF QUESTIONS AND MEASURES FOR SPECIFIC PRACTICE 2.

|  | Questions | Measures |
|---|---|---|
| Q1 | Do you need to purchase new software (e.g., test tools) as a validation environment requirement? | • Needing to purchase new software as a validation environment requirement. |
| Q2 | Do you need to purchase new equipment (e.g., computing or network test environment) as a validation environment requirement? | • Needing to purchase new equipment as a validation environment requirement. |
| Q3 | Do you need new skilled people to operate the new software or equipment? | • Needing of new skilled people to operate the new software or equipment.<br>• # skilled people needed to operate the new software or equipment |
| Q4 | Do you need simulated subsystems or components (by software, electronics, or mechanics) as a validation environment requirement? | • Needing for simulated subsystems or components (by software, electronics, or mechanics) as a validation environment requirement. |

### C. Measures for specific practice 3.

Establish validation procedures and criteria: Establish and maintain procedures and criteria for validation.

A set of questions and measures is presented in the following table related to specific practice 3.

Table 4. set of questions and measures for specific practice 3

|  | Questions | Measures |
|---|---|---|
| Q1 | Do you identify validation procedures to ensure that the products and product components will fulfill its intended use? | • Identifying validation procedures to ensure that the product and product components will fulfill its intended use.<br>• # Products that have a validation procedure.<br>• # Product components for each product that have a validation procedure |
| Q2 | Do you identify validation criteria to ensure that the products and product components will fulfill its intended use? (validation criteria include: product requirements, standards, customer acceptance criteria, environmental performance, the thresholds of performance deviation) | • Identifying validation criteria to ensure that the product and product components will fulfill its intended use.<br>• # Products that have validation criteria.<br>• # Product components for each product that have validation criteria. |
| Q3 | Do you identify the acceptance test cases for the products and product components? | • Identifying the acceptance test cases for the product and product components.<br>  # Products that have acceptance test cases.<br>  # Product components for each product that have acceptance test cases. |
| Q4 | Do you identify the non acceptance test cases for the products and product components? | • Identifying the non acceptance test cases for the product and product components.<br>  # Products that have non acceptance test cases.<br>  # Product components for each product that have non acceptance test cases. |
| Q5 | Do you test and evaluate maintenance, training, and support services? | • Testing and evaluating maintenance, training, and support services |
| Q6 | Do you review the product requirements to ensure that issues affecting validation of products or product components are identified? | • Reviewing the product requirements to ensure that issues affecting validation of product or product components are identified |
| Q7 | Do you document the environment, procedures, input, and output for the validation of the selected products or product components? | • Documenting the environment, procedures, input, and output for the validation of the selected product or product components |

### D. Measures for specific practice 4.

Perform validation: Perform validation on the selected products and product components.

A set of questions and measures is presented in the following table related to specific practice 4.

Table 5 set of questions and measures for specific practice 4.

| | | |
|---|---|---|
| Q1 | Do you validate the products and product components in their operational environment? | • Validating the products in their operational environment.<br>• # Products validated in their operational environment.<br>• # Product components for each product validated in their operational environment. |
| Q2 | Do you validate the products and product components according to the established methods, procedures, and criteria? | • Validating the products and product components according to the established methods, procedures, and criteria. |
| Q3 | Do you collect the resulting data of validation activities of each product or product components? | collecting the resulting data of validation activities of each product or product components |

### E. Measures for specific practice 5.

Analyze validation results: Analyze the results of the validation activities.

A set of questions and measures is presented in the following table related to specific practice 5.

Table 6 set of questions and measures for specific practice5.

| | Questions | Measures |
|---|---|---|
| Q1 | Do you analyze the resulting data against the defined validation criteria and produce analysis reports? (analysis reports indicate whether the needs were met) | • Analyzing the resulting data against the defined validation criteria and produce analysis reports. |
| Q2 | Do you identify the degree of success of failure in the analysis report? | • Identifying the degree of success of failure in the analysis reports. |
| Q3 | Do you compare analysis results with the evaluation criteria? | • Comparing analysis results with the evaluation criteria. |
| Q4 | Do you identify products and product component that do not perform suitably? | • Identifying products and product component that do not perform suitably.<br>• # Products that do not perform suitably.<br>• # Product components that do not perform suitably. |
| Q5 | Do you identify problems with procedures, criteria, and environment? | • Identifying problems with methods, criteria, environment<br>• # Problems with procedures.<br>• # Problems with criteria.<br>• # Problems with environment |

## VI. VALIDITY AND RELIABILITY OF THE DEFINED MEASURES

We have made a questionnaire to prove the reliability and validity of the defined measures and confirm that they are actually measure the five specific practices. The collected data will be analyzed by cronbach alpha reliability in SPSS.

The questionnaire was reviewed and confirmed by academics in software engineering and practitioners in software development in Zarqa University. The questionnaire was filled by system analysts and software engineers. The questionnaire consists of five parts, each part is related to one specific practice of the Validation process, each part consists of a group of statements (measures) related to the specific practice, beside each statement there is five options: strongly agree, agree, neither agree nor disagree, disagree, strongly disagree. The questioner will read the statement and write his opinion of the statement relation with the specific practice by choosing one of the five options, a sample shown in Appendix A.

Cronbach alpha is designed as a measure of internal consistency, that is, do all items measure the same thing? ( measure a single unidimentional structure). Cronbach alpha varies between 0 and 1, the closer the alpha is to 1, the greater the internal consistency of items being assessed [12]. If alpha is less than 0.5 then internal consistency is unacceptable [12]. After applying the collected data on Cronbach Alpha in SPSS we got alpha results between over 0.5 and less than 1.

## VII. CONCLUSION

This paper identified a set of general measures for Validation Process Area (PA) in Capability Maturity Model Integration (CMMI-SW) Staged Representation. The method we used to define the measures is to apply the Goal Question Metrics (GQM) paradigm to the two specific goals and its five specific practices of Validation PA. This work focuses on measurement of a specific process area rather than many process areas at the same time.

The set of measures identified in this paper provide the organization with better insight into the Validation activity, improving the software process towards the goal of having a managed process. The set of measures can be used to control and evaluate software processes and products. Use of the measures varies with the maturity of the process in the organization.

APPENDIX A

# Questionnaire and Analysis

Questionnaire:

This questionnaire is related to the Validation process. Validation demonstrates that products fulfill its intended use when placed in its intended environment.

The Validation process has five goals:
1. Select product for validation.
2. Establish the validation environment.
3. Establish validation procedures and criteria.
4. Perform validation.
5. Analyze validation results.

We would like to measure the achievement of the above goals, so, we define some statements related to each goal. We suppose that the information in these statements help us in achievement of the above five goals.

**Please, fill the enclosed form by writing √ in the suitable place. Responding to this question: do you think that the statements have an effect on the achievement of the goals?**

**1. Goal1: Select product for validation:** Select products and product components to be validated and the validation methods that will be used for each.
 (do you think that these statements have an effect on the achievement of goal1: Select product for validation?)

| statement serial | statements | Strongly agree | Agree | Neither agree nor disagree | disagree | Strongly disagree |
|---|---|---|---|---|---|---|
| 1 | Selecting products and product components to be validated has a positive effect on validation. | | | | | |
| 2 | Selecting the operational behavior for products and product components has a positive effect on validation. | | | | | |
| 3 | Selecting user manual, user interfaces, maintenance services, and training materials to be validated has a positive effect on validation. | | | | | |

33